\documentclass[final,journal,a4paper]{IEEEtran}
\usepackage{amsmath,amsfonts,amssymb}

\newtheorem{thm}{Theorem}
\newtheorem{lem}{Lemma}
\newtheorem{rem}{Remark}

\newcounter{step}
\newenvironment{algorithm}[2]{%
 \begin{list}{%
   \textsf{Step #2\arabic{step}}}%
  {%
   \usecounter{step}%
   \settowidth{\labelwidth}{\textsf{#1}}%
   \addtolength{\labelwidth}{0mm}%
   \setlength{\leftmargin}{\labelwidth}%
   \setlength{\rightmargin}{0pt}%
   \setlength{\labelsep}{0pt}%
   \setlength{\parsep}{0pt}%
   \setlength{\itemsep}{0pt}%
 }}{\end{list}}

\newcommand{\e}{\varepsilon}

\newcommand{\E}{\mathcal{E}}
\newcommand{\I}{\mathcal{I}}
\newcommand{\T}{\mathcal{T}}
\newcommand{\N}{\mathcal{N}}
\newcommand{\U}{\mathcal{U}}
\newcommand{\X}{\mathcal{X}}
\newcommand{\Y}{\mathcal{Y}}

\newcommand{\XX}{\boldsymbol{X}}
\newcommand{\YY}{\boldsymbol{Y}}
\newcommand{\FF}{\boldsymbol{F}}
\newcommand{\CC}{\boldsymbol{C}}
\newcommand{\ba}{\boldsymbol{a}}
\newcommand{\cc}{\boldsymbol{c}}
\newcommand{\ff}{\boldsymbol{f}}
\newcommand{\uu}{\boldsymbol{u}}
\newcommand{\xx}{\boldsymbol{x}}
\newcommand{\yy}{\boldsymbol{y}}
\newcommand{\PPsi}{\boldsymbol{\Psi}}
\newcommand{\pphi}{\boldsymbol{\phi}}
\newcommand{\ppsi}{\boldsymbol{\psi}}

\newcommand{\hc}{\widehat{c}}
\newcommand{\hphi}{\widehat{\phi}}

\newcommand{\hcc}{\widehat{\cc}}
\newcommand{\hxx}{\widehat{\xx}}
\newcommand{\hpphi}{\widehat{\pphi}}

\newcommand{\lrB}[1]{\left[{#1}\right]}
\newcommand{\lrb}[1]{\left\{{#1}\right\}}
\newcommand{\lrsb}[1]{\left({#1}\right)}
\newcommand{\lrbar}[1]{\left|{#1}\right|}

\newcommand{\fMAP}{\ff_{\mathrm{MAP}}}
\newcommand{\Prob}{\mathrm{Prob}}

\allowdisplaybreaks

\begin{document}
\title{
 Successive-Cancellation Decoding\\
 of Linear Source Code
}
\author{Jun~Muramatsu\\
NTT Communication Science Laboratories, NTT Corporation\\
Hikaridai 2-4, Seika-cho, Soraku-gun, Kyoto 619-0237, Japan}
\maketitle

\begin{abstract}
 This paper investigates the error probability
 of several decoding methods for a source code
 with decoder side information,
 where the decoding methods are:
 1) symbol-wise maximum a posteriori decoding,
 2) successive-cancellation decoding, and
 3) stochastic successive-cancellation decoding.
 The proof of the effectiveness of a decoding method
 is reduced to that for an arbitrary decoding method,
 where `effective' means that the error probability
 goes to zero as $n$ goes to infinity.
 Furthermore, we revisit the polar source code
 showing that stochastic successive-cancellation decoding,
 as well as successive-cancellation decoding,
 is effective for this code.
\end{abstract}
  
\section{Introduction}

Successive-cancellation (SC) decoding is one of the elements
constituting the polar code introduced by Ar\i{}kan~\cite{A09}.
This paper investigates the error probability of SC decoding
for a source code with decoder side information
by extending the results in~\cite{A10,S12} to general linear source codes.
It is shown that if for a given encoder there is a decoder
such that the block error probability is $o(1/n)$,
then the block error probability of an SC decoder
for the same encoder is $o(1)$.
Furthermore, we introduce stochastic successive-cancellation (SSC)
decoding and show that it is equivalent to the constrained-random-number
generator introduced in~\cite{CRNG}.
It is shown that if for a given encoder there is a decoder
such that the block error probability is $o(1)$,
then the block error probability of an SC decoder for the same
encoder is $o(1)$.
It is also shown that the error probability of the symbol-wise maximum a
posteriori decoding of a linear source code and
the SSC decoder of the polar source code goes to zero
as the block length goes to infinity.

It should be noted that the results of this paper
can be applied to the channel coding as introduced in~\cite{A10,SW2CC,S12}.
In particular, the syndrome decoding
is the case when a channel is additive,
a parity check matrix corresponds to a source encoding function,
the syndrome of a channel output corresponds to a codeword of the source
code without decoder side information,
and the kernel of the parity check matrix forms the channel inputs,
that is, the codewords for a channel code.

Throughout the paper, we use the following notations.
For random variable $U$,
let $\U$ be the alphabet of $U$,
$\mu_{U}$ be the distribution of $U$,
and $\mu_{U|V}$ be the conditional distribution of $U$ for
a given random variable $V$.
Let $H(U|V)$ be the conditional entropy of $U$ for a given $V$,
where we assume that the base of $\log$ is
the cardinality $|\U|$ of $\U$.
A column vector is denoted by a boldface letter $\uu$,
where its dimension depends on the context.
We define $u_i^j\equiv(u_i,\ldots,u_j)$,
where $u_i^j$ is the null string when $i>j$.
Let $\chi(\cdot)$ be a support function defined as
\[
 \chi(\mathrm{S})
 \equiv
 \begin{cases}
  1,&\text{if the statement $\mathrm{S}$ is true}
  \\
  0,&\text{if the statement $\mathrm{S}$ is false}.
 \end{cases}
\]

\section{Symbol-wise Maximum A Posteriori Decoding}

First, we revisit symbol-wise maximum a posteriori (SMAP) decoding,
which is used for the conventional decoding
of a low density parity check code.
Although the symbol error rate 
(the Hamming distance between a source output and its reproduction
 divided by the block length $n$)
is discussed with symbol-wise maximum a posteriori decoding,
we focus on the block error probability
(an error occurs when a source output and its reproduction are different,
 that is, the Hamming distance is positive)
throughout this paper.

Let $(A,\pphi)$ be a pair consisting of a source encoder $A:\X^n\to\X^l$
and a decoder $\pphi:\X^l\times\Y^n\to\X^n$ with side information.
Let $\cc_1\equiv A\xx$ be the codeword of a source output $\xx\in\X^n$.
The decoder $\hpphi\equiv\{\hphi_i\}_{i=1}^n$
is constructed by using functions reproducing the $i$-th coordinate as
\[
 \hphi_i(\cc_1,\yy)
 \equiv
 \arg\max_{x_i}\mu_{X_i|\CC_1\YY}(x_i|\cc_1,\yy).
\]
It should be noted that
when $(\XX,\YY)\equiv(X^n,Y^n)$ is memoryless and $A$ is a sparse matrix
we can use the sum-product algorithm to obtain an approximation of
$\mu_{X_i|\CC_1\YY}(x_i|\cc_1,\yy)$.

We have the following theorem.
\begin{thm}
\label{thm:smap}
The error probability of the code $(A,\hpphi)$ is bounded as
\[
 \Prob(\hpphi(A\XX,\YY)\neq\XX)
 \leq
 n\Prob(\pphi(A\XX,\YY)\neq\XX),
\]
where the right hand side of this inequality
goes to zero as $n\to\infty$ when $\Prob(\pphi(A\XX,\YY)\neq\XX)=o(1/n)$.
\end{thm}
\begin{IEEEproof}
 Let 
 $\phi_i(\cc_1,\yy)$ be the $i$-th coordinate of $\pphi(\cc_1,\yy)$.
 Then we have
 \begin{align}
  &
  \Prob(\hpphi(A\XX,\YY)\neq \XX)
  \notag
  \\*
  &=
  \Prob(\hphi_i(A\XX,\YY)\neq X_i\ \text{for some}\ i)
  \notag
  \\*
  &\leq
  \sum_{i=1}^n\Prob(\hphi_i(A\XX,\YY)\neq X_i)
  \notag
  \\
  &=
  \sum_{i=1}^n
  \Prob(\arg\max_{x_i}\mu_{X_i|\CC_1\YY}(x_i|A\XX,\YY)\neq X_i)
  \notag
  \\
  &\leq
  \sum_{i=1}^n
  \Prob(\phi_i(A\XX,\YY)\neq X_i)
  \notag
  \\
  &\leq
  \sum_{i=1}^n
  \Prob(\pphi(A\XX,\YY)\neq \XX)
  \notag
  \\
  &\leq
  n\Prob(\pphi(A\XX,\YY)\neq \XX),
 \end{align}
 where the first inequality comes from the union bound,
 the second inequality comes from the fact that
 the maximum a posteriori decision minimizes the error probability,
 and the third inequality comes from the fact that
 $\phi_i(\cc_1,\yy)\neq x_i$ implies $\pphi(\cc_1,\yy)\neq\xx$.
\end{IEEEproof}

It is known that, when $l/n>H(X^n|Y^n)/n$
there is an encoding function $A:\X^n\to\X^l$
such that error probability $\Prob(\pphi(A\XX,\YY)\neq\XX)$
is close to zero for all sufficiently large $n$~\cite{CSI82,SWLDPC},
where we can use one of the following decoders:
\begin{itemize}
 \item
 the typical set decoder defined as
 \[
  \pphi(\cc_1,\yy)
  \equiv
  \begin{cases}
   \hxx
   &\text{if there is a unique}\ \hxx\in\T_{\XX|\YY,\e}(\yy)
   \\
   \text{`error'}
   &\text{otherwise},
  \end{cases}
 \]
 where
 \[
  \T_{\XX|\YY,\e}(\yy)
  \equiv
  \lrb{\xx:
   |\log\mu_{\XX|\YY}(\xx|\yy)-H(\XX|\YY)|\leq n\e
  }
 \]
 is a conditional typical set,
 \item
 the maximum a posteriori probability decoder\footnote{The right
   hand side of the third equality of (\ref{eq:map})
   might be called the maximum-likelihood decoder.}
 defined as
 \begin{align}
  \pphi(\cc_1,\yy)
  &\equiv
  \arg\max_{\xx}\mu_{\XX|\CC_1\YY}(\xx|\cc_1,\yy)
  \notag
  \\
  &=
  \arg\max_{\xx}\mu_{\XX\CC_1\YY}(\xx,\cc_1,\yy)
  \notag
  \\
  &=
  \arg\max_{\xx:A\xx=\cc_1}\mu_{\XX\YY}(\xx,\yy)
  \notag
  \\
  &=
  \arg\max_{\xx:A\xx=\cc_1}\mu_{\XX|\YY}(\xx|\yy),
  \label{eq:map}
 \end{align}
 where the third equality comes from the fact that
 $\mu_{\XX\CC_1\YY}(\xx,\cc_1,\yy)=\mu_{\XX\YY}(\xx,\yy)$
 when $A\xx=\cc_1$
 and $\mu_{\XX\CC_1\YY}(\xx,\cc_1,\yy)=0$ when $A\xx\neq\cc_1$.
\end{itemize}

The following sections
show upper bounds of the error probability for several decoders
in terms of the error probability of a code $(A,\pphi)$,
where $\pphi$ is an arbitrary decoder.
It should be noted that we can use one of the decoders mentioned above.
We can reduce the effectiveness of the decoders
to that of an arbitrary decoder,
where `effective' means that the error probability goes to zero
as $n$ goes to infinity.
For example, \cite{SDECODING,SW2CC} show that
a decoder using a constrained-random-number generator is effective
by showing that the maximum a posteriori probability decoder is effective.

\section{Decoding Extended Codeword}

Let $A:\X^n\to\X^l$ be an encoder of a source code
with decoder side information.
Here, we assume that, for a given $A$
there is a function $B:\X^n\to\X^{n-l}$
and a bijection $Q:\X^n\to\X^n$ such that
\begin{equation}
 Q(A\xx,B\xx)=\xx\quad\text{for all}\ \xx\in\X^n.
 \label{eq:Q}
\end{equation}
In particular, this condition is satisfied when $A$ is a full-rank matrix.
We define the bijection $[A,B]:\X^n\to\X^n$ as
$[A,B]\xx \equiv (A\xx,B\xx)$.

Let $\I_0$ and $\I_1$ be a partition of $\N\equiv\{1,\ldots,n\}$,
that is, they satisfy $I_0\cap\I_1=\emptyset$ and $\I_0\cup\I_1=\N$.
We call $\I_0$ and $\I_1$ {\em ordered} when
$\I_1=\{1,\ldots,l\}$ and $\I_0=\{l+1,\ldots,n\}$.
For a vector $\cc\equiv(c_1,\ldots,c_n)\in\X^n$,
define $\cc_0\in\X^{n-l}$ and $\cc_1\in\X^l$
so that $c_i$ is a symbol in $\cc_b$ when $i\in\I_b$ for every
$b\in\{0,1\}$.
In the following, we assume that $A\xx=\cc_1$ and $B\xx=\cc_0$,
where corresponding index sets $\I_1$ and $\I_0$ may not be ordered
in the bijection $[A,B]$.
We call $(\cc_0,\cc_1)$ the {\em extended codeword} of $\cc_1$.
In the following, we denote $\cc=(\cc_0,\cc_1)$ omitting
the dependence on $(\I_0,\I_1)$.

Let $\ff:\X^l\times\Y^n\to\X^n$ be a function
that reproduces the extended codeword by using the side information.
For a codeword $\cc_1\in\X^l$ and side information $\yy\in\Y^n$,
the source decoder $\ppsi$ with side information is defined as
\begin{equation}
 \ppsi(\cc_1,\yy)\equiv Q(\ff(\cc_1,\yy)).
 \label{eq:exc}
\end{equation}

In the context of the polar source codes,
$\cc_0$ corresponds to unfrozen symbols
and $Q$ corresponds to the final step of SC decoding.
We have the following lemma for a general case.
\begin{lem}
\label{lem:exc}
Let $\CC_1\equiv A\XX$ and $\CC_0\equiv B\XX$. Then we have
\[
 \Prob(\ppsi(A\XX,\YY)\neq \XX)=\Prob(\ff(\CC_1,\YY)\neq (\CC_0,\CC_1)).
\]
\end{lem}
\begin{IEEEproof}
 We have
 \begin{align}
 &
 \Prob(\ppsi(A\XX,\YY)\neq \XX)
 \notag
 \\*
 &=
 \sum_{\xx,\yy}\mu_{\XX\YY}(\xx,\yy)\chi(\ppsi(A\xx,\yy)\neq \xx)
 \notag
 \\
 &=
 \sum_{\xx,\yy,\cc_0,\cc_1}\mu_{\XX\YY}(\xx,\yy)\chi(A\xx=\cc_1)\chi(B\xx=\cc_0)
 \notag
 \\*
 &\quad\cdot
 \chi(\ppsi(\cc_1,\yy)\neq \xx)
 \notag
 \\
 &=
 \sum_{\xx,\yy,\cc_0,\cc_1}\mu_{\XX\YY}(\xx,\yy)\chi(A\xx=\cc_1)\chi(B\xx=\cc_0)
 \notag
 \\*
 &\quad\cdot
 \chi(\ff(\cc_1,\yy)\neq Q^{-1}(\xx))
 \notag
 \\
 &=
 \sum_{\xx,\yy,\cc_0,\cc_1}\mu_{\XX\YY}(\xx,\yy)\chi(A\xx=\cc_1)\chi(B\xx=\cc_0)
 \notag
 \\*
 &\quad\cdot
 \chi(\ff(\cc_1,\yy)\neq (A\xx,B\xx))
 \notag
 \\
 &=
 \sum_{\xx,\yy,\cc_0,\cc_1}\mu_{\XX\YY}(\xx,\yy)\chi(A\xx=\cc_1)\chi(B\xx=\cc_0)
 \notag
 \\*
 &\quad\cdot
 \chi(\ff(\cc_1,\yy)\neq (\cc_0,\cc_1))
 \notag
 \\
 &=
 \sum_{\cc_0,\cc_1,\yy}\mu_{\CC_0\CC_1\YY}(\cc_0,\cc_1,\yy)
 \chi(\ff(\cc_1,\yy)\neq (\cc_0,\cc_1)),
 \notag
 \\
 &=
 \Prob(\ff(\CC_1,\YY)\neq (\CC_0,\CC_1)),
\end{align}
where the third equality comes from the fact that $Q$ is bijective,
and in the sixth equality we define
\begin{equation}
 \mu_{\CC_0\CC_1\YY}(\cc_0,\cc_1,\yy)\equiv\mu_{\XX\YY}(Q(\cc_1,\cc_0),\yy)
 \label{eq:joint-exc}
\end{equation}
and use the fact that for all $\cc_0$ and $\cc_1$
there is a unique $\xx$ satisfying $A\xx=\cc_1$ and $B\xx=\cc_0$.
\end{IEEEproof}

In the following, we investigate the decoding error probability
for an extended codeword.

\section{Successive-Cancellation Decoding}

This section investigates the error probability
of the (deterministic) SC decoding.
For a source encoder $A:\X^n\to\X^l$,
let $B$, $Q$, $\CC_0$, and $\CC_1$ be defined as in the previous section.

For a codeword $\cc_1\in\X^l$ and side information $\yy\in\Y^n$,
the output $\hcc\equiv \ff(\cc_1,\yy)$ of an SC decoder $\ff$ is defined
recursively as
\[
 \hc_i
 \equiv
 \begin{cases}
  f_i(\hc_1^{i-1},\yy)
  &\text{if}\ i\in\I_0
  \\
  c_i
  &\text{if}\ i\in\I_1
 \end{cases}
\]
by using functions $\{f_i\}_{i\in\I_0}$ defined as
\begin{equation}
 f_i(c_1^{i-1},\yy)
 \equiv
 \arg\max_{c_i}\mu_{C_i|C_1^{i-1}\YY}(c_i|c_1^{i-1},\yy),
 \label{eq:sc-map}
\end{equation}
which is known as the maximum a posteriori decision rule, where
$\mu_{C_i|C_1^{i-1}\YY}$ is the conditional probability defined as
\begin{equation}
 \mu_{C_i|C_1^{i-1}\YY}(c_i|c_1^{i-1},\yy)
 \equiv
 \frac{\sum_{c_{i+1}^n}\mu_{\CC_0\CC_1\YY}(\cc_0,\cc_1,\yy)}
 {\sum_{c_i^n}\mu_{\CC_0\CC_1\YY}(\cc_0,\cc_1,\yy)}
 \label{eq:muCgCY}
\end{equation}
by using $\mu_{\CC_0\CC_1\YY}$ defined by (\ref{eq:joint-exc}).

To simplify the notation,
we define $f_i(\hc_1^{i-1},\yy)\equiv c_i$ when $i\in\I_1$
although $c_i$ does not depend on $\hc_1^{i-1}$ and $\yy$.
We have the following lemma.
\begin{lem}
\label{lem:sc}
\begin{align*}
 &
 \Prob(\ff(\CC_1,\YY)\neq (\CC_0,\CC_1))
 \notag
 \\*
 &\leq
 \sum_{i\in\I_0}
 \Prob(f_i(\CC_1^{i-1},\YY)\neq C_i).
\end{align*}
\end{lem}
\begin{IEEEproof}
As with the proof in~\cite{A09}, we can express the block error events
$\ff(\cc_1,\yy)\neq (\cc_0,\cc_1)$ as $\E\equiv\bigcup_{i=1}^n\E_i$,
where
\[
 \E_i
 \equiv
 \lrb{
  (\cc,\yy):
  \begin{aligned}
   &f_j(c_1^{j-1},\yy)=c_j\ \text{for all}\ j\in\{1,\ldots,i-1\}
   \\
   &f_i(c_1^{i-1},\yy)\neq c_i
  \end{aligned}
 }
\]
is an event where the first decision error in SC decoding occurs
at stage $i$.
The decoding error probability for a extended codeword is evaluated as
\begin{align}
 &
 \Prob(\ff(\CC_1,\YY)\neq (\CC_0,\CC_1))
 \notag
 \\*
 &=
 \Prob((\CC_0,\CC_1,\YY)\in\E)
 \notag
 \\
 &\leq
 \sum_{i=1}^n
 \Prob((\CC_0,\CC_1,\YY)\in\E_i)
 \notag
 \\
 &=
 \sum_{i\in\I_0}
 \Prob((\CC_0,\CC_1,\YY)\in\E_i),
 \notag
 \\
 &\leq
 \sum_{i\in\I_0}
 \Prob(f_i(\CC_1^{i-1},\YY)\neq C_i),
\end{align}
where the first inequality comes from the union bound,
the second equality comes from the fact that
$f_i(\CC_1^{i-1},\YY)=C_i$ when $i\in\I_1$,
and the last inequality comes from the fact that
$(\cc_0,\cc_1,\yy)\in\E_i$ implies $f_i(c_1^{i-1},\yy)\neq c_i$.
\end{IEEEproof}

When the index sets $\I_1$ and $\I_0$ are not ordered
like the polar source codes~\cite{A10,S12},
$f_i$ defined by (\ref{eq:sc-map}) may not use
the full information of a codeword $\cc_1\equiv\{c_i\}_{i\in\I_1}$.
Borrowing words from~\cite{A09},
$f_i$ treats {\em future symbols} as random variables
rather than as known symbols.
In other words, $f_i$ ignores the future symbols in a codeword $\cc_1$.
This implies that $\{f_i\}_{i=1}^n$ is different from
the optimum maximum a posteriori decoder defined as
\[
 \fMAP(\cc_1,\yy)\equiv
 \arg\max_{\cc_0}\mu_{\CC_0|\CC_1\YY}(\cc_0|\cc_1,\yy).
\]

The following investigates
the error probability of the (deterministic) SC decoding
by assuming that the index sets $\I_1$ and $\I_0$ are ordered,
that is, $\I_1=\{1,\ldots,l\}$ and $\I_0=\{l+1,\ldots,n\}$.
This implies that for every $i\in\I_0$,
$f_i$ defined by (\ref{eq:sc-map}) uses the full information
of a codeword $\{c_i\}_{i\in\I_1}$.
\begin{lem}
\label{lem:error-ssc}
For a source encoder $A:\X^n\to\X^l$ and
decoder $\pphi:\X^l\times\Y^n\to\X^n$
with side information, 
let $B$, $Q$, $\CC_0$, and $\CC_1$
be as defined in the previous section,
where it is assumed that the index sets $\I_1$ and $\I_0$ are ordered.
Then we have
\[
 \Prob(f_i(C_1^{i-1},\YY)\neq C_i)
 \leq
 \Prob(\pphi(A\XX,\YY)\neq\XX)
\]
 for all $i\in\I_0$.
\end{lem}
\begin{IEEEproof}
 For $i\in\I_0$,
 let $f'_i(\cc_1,\yy)$ be the $i$-th coordinate
 of the extended codeword of $Q^{-1}(\pphi(\cc_1,\yy))$.
 Then we have the fact that
 \begin{align}
  f'_i(\cc_1,\yy)\neq c_i
  &\Rightarrow Q^{-1}(\pphi(\cc_1,\yy))\neq (\cc_0,\cc_1)
  \notag
  \\
  &\Leftrightarrow \pphi(\cc_1,\yy)\neq Q(\cc_1,\cc_0)
  \notag
  \\
  &\Leftrightarrow \pphi(A\xx,\yy)\neq \xx
  \label{eq:sc-error}
 \end{align}
 for all $\xx$ satisfying $A\xx=\cc_1$ and $B\xx=\cc_0$,
 where the second equivalence comes from the fact that $Q$ is bijective,
 and the third equivalence comes from (\ref{eq:Q}).
 Then we have
 \begin{align}
  &
  \Prob(f_i(C_1^{i-1},\YY)\neq C_i)
  \notag
  \\*
  &=
  \Prob(\arg\max_{c_i}\mu_{C_i|C_1^{i-1}\YY}(c_i|C_1^{i-1},\YY)\neq C_i)
  \notag
  \\
  &\leq
  \Prob(\arg\max_{c_i}\mu_{C_i|\CC_1\YY}(c_i|\CC_1,\YY)\neq C_i)
  \notag
  \\
  &\leq
  \Prob(f'_i(\CC_1,\YY)\neq C_i)
  \notag
  \\
  &\leq
  \Prob(\pphi(A\XX,\YY)\neq \XX),
 \end{align}
 where the first inequality comes from
 Lemma~\ref{lem:mapUgVW} in the Appendix
 and the fact that $\CC_1=C_1^l$,
 the second inequality comes from the fact that
 the maximum a posteriori decision rule
 minimizes the decision error probability,
 and the last inequality comes from (\ref{eq:sc-error}).
\end{IEEEproof}

From Lemmas~\ref{lem:exc}--\ref{lem:error-ssc}
and the fact that $|\I_0|\leq n$,
we have the following theorem,
which implies that SC decoding is effective
when for a given encoding function $A$
there is an effective decoding function $\pphi$.
\begin{thm}
\label{thm:sc}
For a source code $(A,\pphi)$
with decoder side information,
error probability of the (deterministic) SC decoding $\ppsi$
is bounded as
\[
 \Prob(\ppsi(A\XX,\YY)\neq\XX)
 \leq 
 n\Prob(\pphi(A\XX,\YY)\neq\XX),
\]
where the right hand side of this inequality
goes to zero as $n\to\infty$ when $\Prob(\pphi(A\XX,\YY)\neq\XX)=o(1/n)$.
\end{thm}

It should be noted again that the
index sets $\I_1$ and $\I_0$ are ordered,
while they are not ordered in the original polar source code.
In contrast, we can use an arbitrary function $B$ that satisfies
the assumption and rearrange the index sets $\I_1$ and $\I_0$
so that they are ordered,
while they are fixed in the original polar source code.

\section{Stochastic Successive-Cancellation Decoding}

This section introduces
stochastic successive-cancellation (SSC) decoding,
which is known as randomized rounding in the context of polar codes.

When $i\in\I_0$, we replace $f_i$ defined in (\ref{eq:sc-map})
by the stochastic decision rule generating $c_i$ randomly
subject to the probability distribution
$\{\mu_{C_i|C_1^{i-1}\YY}(c'_i|c_1^{i-1},\yy)\}_{c'_i\in\X}$
for a given $(c_1^{i-1},\yy)$.
Let $F_i$ be the stochastic decision rule described above.
Let $\FF$ be the stochastic decoder by using $F_i$ instead of $f_i$
when $i\in\I_0$.
We denote the stochastic decoder corresponding to (\ref{eq:exc})
by $\PPsi$.
An analysis of the error probability will be presented
in the next section.

\section{Implementation of Successive-Cancellation Decoding}

In this section,
we assume that $A$ is a full-rank $l\times n$ (sparse) matrix.
Without loss of generality, we can assume that
the right part of $A$ is an invertible $l\times l$ matrix.
This condition is satisfied for an arbitrary
full-rank matrix $A$ by using a permutation matrix
$S$, where $AS$ satisfies the condition,
and the codeword can be obtained as $A\xx=AS[S^{-1}\xx]$.

Let $B$ be an $[n-l]\times n$ matrix,
where the left part of $B$ is an invertible $[n-l]\times[n-l]$ matrix.
Then we have the fact that
by concatenating row vectors of $B$ to $A$,
we obtain the invertible $n\times n$ matrix $[A,B]$,
that is, $[A,B]$ is bijective.
By using $A$ and $B$,
we can construct a successive-cancellation decoder
that reproduces an extended codeword
with $\I_1=\{1,\ldots,l\}$ and $\I_0=\{l+1,\ldots,n\}$.

Here, let us assume that
the left part of $B$ is the $[n-l]\times[n-l]$ identity matrix
and the right part of $B$ is the $[n-l]\times l$ zero matrix.
It should be noted that
a similar discussion is possible
when the identity matrix is replaced by a permutation matrix.

Since the left part of $B$ is the
$[n-l]\times[n-l]$ identity matrix,
then, for all $i\in\{0,\ldots,l-1\}$,
the $(i,i)$-element of $[A,B]$ is $1$,
which is the only positive element in $i$-th row of $[A,B]$.
Then we have the fact that
\[
 C_{l+j}=X_j\quad\text{for all}\ j\in\{1,\ldots,n-l\},
\]
which implies $\CC_0=X_1^{n-l}$.

First, we reduce the conditional probability 
$\mu_{C_i|C_1^{i-1}\YY}(c_i|c_1^{i-1},\yy)$
defined by (\ref{eq:muCgCY}).
For $i\in\{l+1,\ldots,n\}$ and $j\equiv i-l$, we have
\begin{align}
 \mu_{C_i|C_1^{i-1}\YY}(c_i|c_1^{i-1},\yy)
 &=
 \frac{\mu_{C_1^1\YY}(c_1^i,\yy)}
 {\mu_{C_1^{i-1}\YY}(c_1^{i-1},\yy)}
 \notag
 \\
 &=
 \frac{\mu_{C_1^lC_{l+1}^{l+j}\YY}(c_1^l,c_{l+1}^{l+j},\yy)}
  {\mu_{C_1^lC_{l+1}^{l+j-1}\YY}(c_1^l,c_{l+1}^{l+j-1},\yy)}
 \notag
 \\
 &=
 \frac{\mu_{C_{l+1}^{l+j}\CC_1,\YY}(c_{l+1}^{l+j},\cc_1,\yy)}
  {\mu_{C_{l+1}^{l+j-1}\CC_1\YY}(c_{l+1}^{l+j-1},\cc_1,\yy)}
 \notag
 \\
 &=
 \frac{\mu_{X_1^j\CC_1,\YY}(c_{l+1}^{l+j},\cc_1,\yy)}
  {\mu_{X_1^{j-1}\CC_1\YY}(c_{l+1}^{l+j-1},\cc_1,\yy)}
 \notag
 \\
 &=
 \mu_{X_j|X_1^{j-1}\CC_1,\YY}(c_{l+j}|c_{l+1}^{l+j-1},\cc_1,\yy),
\end{align}
where the third equality comes from the fact that $\I_1=\{1,\ldots,l\}$
and the fourth equality comes from Lemma~\ref{lem:U=U'} in the Appendix
and the fact that $C_{l+j}=X_j$ for all $j\in\{1,\ldots,n-l\}$.
By substituting $c_{l+1}^i=x_1^j$, we have
\begin{align}
 \mu_{C_i|C_1^{i-1}\YY}(x_j|x_1^{j-1},\yy)
 &=
 \mu_{X_j|X_1^{j-1}\CC_1\YY}(x_j|x_1^{j-1},\cc_1,\yy)
 \notag
 \\
 &=
 \frac{\sum_{x_{j+1}^n}\mu_{\XX|\YY}(\xx|\yy)\chi(A\xx=\cc_1)}
 {\sum_{x_j^n}\mu_{\XX|\YY}(\xx|\yy)\chi(A\xx=\cc_1)}
 \label{eq:crng}
\end{align}
for $i\in\{l+1,\ldots,n\}$ and $j\equiv i-l$.
It should be noted that the right hand side of the second equality
appears in the constrained-random-number generation
algorithm~\cite[Eq.~(41)]{CRNG}\footnote{In~\cite[Eq.~(41)]{CRNG},
$\mu_{X_j|X_1^{j-1}}$ should be replaced by $\mu_{X_j|X_1^{j-1}\YY}$.}.
This implies that the constrained-random-number generator
can be considered as an SSC decoding $\PPsi$ of the extended
codeword specified in the previous section,
where we have assumed that 
this algorithm uses the full information of the codeword $c_1^l$
for every $i\in\{l+1,\ldots,n\}$.

Next, we assume that $(X^n,Y^n)$ is memoryless
and reduce the condition $A\xx=\cc_1$ to improve the algorithm.
This idea has already been presented in~\cite{CRNG-VLOSSY}.
Let $\ba_j$ be the $j$-th column vector of $A$.
Let $A_1^{j-1}$ be the sub-matrix of $A$
obtained by using $\{\ba_{j'}\}_{j'=1}^{j-1}$
and $A_j^n$ be that obtained by using $\{\ba_{j'}\}_{j'=j}^n$.
At the computation of (\ref{eq:crng}) for $j\in\{1,\ldots,n-l\}$,
we can assume that $x_1^{j-1}$ has already been determined.
Furthermore, we have the fact that the condition $A\xx=\cc_1$
is equivalent to $A_j^nx_j^n=\cc_1-A_1^{j-1}x_1^{j-1}$.
Then, 
by letting $\cc'_1(j)\equiv\cc_1-A_1^{j-1}x_1^{j-1}$,
we can reduce (\ref{eq:crng}) as follows:
\begin{align}
 &\mu_{C_i|C_1^{i-1}\YY}(x_j|x_1^{j-1},\yy)
 \notag
 \\*
 &=
 \frac{\sum_{x_{j+1}^n}\mu_{\XX|\YY}(\xx|\yy)\chi(A\xx=\cc_1)}
 {\sum_{x_j^n}\mu_{\XX|\YY}(\xx|\yy)\chi(A\xx=\cc_1)}
 \notag
 \\
 &=
 \frac{
 \sum_{x_{j+1}^n}
 \lrB{\prod_{k=j}^n
 \mu_{X_k|Y_k}(x_k|y_k)}
 \chi(A_j^nx_j^n=\cc'_1(j))}
{\sum_{x_j^n}
 \lrB{\prod_{k=j}^n
 \mu_{X_k|Y_k}(x_k|y_k)}
 \chi(A_j^nx_j^n=\cc'_1(j))}.
\label{eq:crng-sp}
\end{align}
It should be noted that we can obtain $A_j^n$ recursively by deleting
the left-end column vector of $A_{j-1}^n$.
We can obtain the vector $\cc'_1(j)$
recursively by using the relations
\begin{align*}
 \cc'_1(1)
 &\equiv \cc_1
 \\
 \cc'_1(j)
 &\equiv \cc'_1(j-1)-x_{j-1}\ba_{j-1}\ \text{for}\ j\in\{2,\ldots,n-l\}.
\end{align*}
These operations reduce the computational complexity of the algorithm.
It should also be noted that
the sum-product algorithm is available
for the approximate computation of (\ref{eq:crng-sp})
when $A$ is a sparse matrix.

Next, we convert the reproduction of a extended codeword
to the reproduction of a source output.
When $j=n-l$, we have obtained the extended codeword $(\cc_1,\cc_0)$,
where $\cc_0\equiv c_{l+1}^n=x_1^{n-l}$.
We can reproduce the source output $\xx$
by using the relation $\xx\equiv [A,B]^{-1}\cc$,
where $[A,B]^{-1}$ is the inverse of the concatenation of $A$ and $B$.
Then we have the relations
\begin{align*}
 \cc_1 &= A_1^{n-l}x_1^{n-l} + A_{n-l+1}^nx_{n-l+1}^n
 \\
 \cc_0 &=x_1^{n-l}
\end{align*}
from the assumptions of $A$ and $B$.
Since 
\[
 \cc'_1(n-l+1)=\cc_1-A_1^{n-l}x_1^{n-l},
\]
we obtain $x_{n-l+1}^n$ as
\[
 x_{n-l+1}^n= [A_{n-l+1}^n]^{-1}\cc'_1(n-l+1),
\]
where $[A_{n-l+1}^n]^{-1}$ is the inverse of $A_{n-l+1}^n$.

Finally, we summarize the decoding algorithm.
We assume that $(X^n,Y^n)$ is memoryless,
$A$ is an $l\times n$ (sparse) matrix satisfying that
$A_{n-l+1}^n$ is an $l\times l$ invertible matrix,
and $B$ is an $[n-l]\times n$ matrix satisfying that $B_1^{n-l}$ is an
$[n-l]\times[n-l]$ identity matrix.

\noindent{\bf SC/SSC Decoding Algorithm Using Sum-Product Algorithm:}

\begin{algorithm}{Step 99}{}
 \item
 Let $j\leftarrow1$ and $\cc'_1\leftarrow\cc_1$.
 \item
 Calculate the conditional probability distribution
 $\mu_{C_{l+j}|C_1^{l+j-1}\YY}$ as
 \begin{align}
  &
  \mu_{C_{l+j}|C_1^{l+j-1}\YY}(c_j|c_1^{j-1},\yy)
  \notag
  \\*
  &=
  \mu_{X_j|X_1^{j-1}\CC_1\YY}(x_j|x_1^{j-1},\cc_1,\yy)
  \notag
  \\*
  &=
  \frac{\displaystyle
   \sum_{x_{j+1}^n}
   \lrB{\prod_{k=j}^n\mu_{X_k|Y_k}(x_k|y_k)}
   \chi(A_j^nx_j^n=\cc'_1)
  }{\displaystyle
   \sum_{x_j^n}
   \lrB{\prod_{k=j}^n\mu_{X_k|Y_k}(x_k|y_k)}
   \chi(A_j^nx_j^n=\cc'_1)
  }
  \label{eq:sum-product}
 \end{align}
 by using $x_1^{j-1}$, $y_{j+1}^n$, $A_j^n$, and $\cc'_1$,
 where we define $c_{l+1}^{l+j}\equiv x_1^j$.
 It should be noted that the sum-product algorithm can be employed
 to obtain an approximation of (\ref{eq:sum-product}).
 \item
 For the deterministic SC decoding,
 let $x_j$ be defined as
 \[
  x_j\equiv\arg\max_{x'_j}\mu_{C_{l+j}|C_1^{l+j-1}\YY}(x'_j|x_1^{j-1},\yy).
 \]
 For the SSC decoding,
 generate and record a random number $x_j$ subject to the
 distribution
 $\{\mu_{C_{l+j}|C_1^{l+j-1}\YY}(x'_j|x_1^{j-1},\yy)\}_{x'_j\in\X}$.
 \item
 Let $\cc'_1\leftarrow \cc'_1-x_j\ba_j$.
 \item
 If $j=n-l$, then compute $x_{l+1}^n\equiv [A_{n-l+1}^n]^{-1}\cc'_1$,
 output $x_1^n$ and terminate.
 \item
  Let $j\leftarrow j+1$ and go to {\sf Step 2}.
\end{algorithm}

Since the SSC decoder is equivalent to
a constrained-random-number generator generating a random sequence
subject to the a posteriori probability distribution
$\mu_{\XX|\CC_1\YY}$ \cite[Theorem 5]{CRNG},
we have the following theorem from the fact that
the error probability of a stochastic decision
with an a posteriori probability distribution is at most twice
that of {\em any} decision rule~\cite[Lemma 3]{SDECODING}.
\begin{thm}
\label{thm:ssc}
For a linear source code $(A,\phi)$ with decoder side information,
the decoding error of the SSC decoding algorithm
is bounded as
\[
 \Prob(\PPsi(A\XX,\YY)\neq\XX)
 \leq
 2\Prob(\pphi(A\XX,\YY)\neq\XX),
\]
where the right hand side of this inequality
goes to zero as $n\to\infty$ when $\Prob(\pphi(A\XX,\YY)\neq\XX)=o(1)$.
\end{thm}

\section{Analysis When Index Sets Are Not Ordered}
In the previous sections, it was assumed that
the index sets $\I_1$ and $\I_0$
corresponding to $\cc_1=A\xx$ and $\cc_0=B\xx$ are ordered,
that is, $\I_1=\{1,\ldots,l\}$ and $\I_0=\{l+1,\ldots,n\}$.
This section investigates the case when they are not ordered.
The following lemma asserts that
the effectiveness of the decoder is reduced to a condition
where the sum of the conditional entropies
corresponding to the complement of the codeword
goes to zero as $n\to\infty$.
\begin{lem}
\label{lem:error<H}
Let $\ppsi$ and $\PPsi$ be the SC and SSC decoding functions,
respectively.
Then
\begin{align*}
 \Prob(\ppsi(A\XX,\YY)\neq \XX)
 &\leq
 \frac1{2\log 2}
 \sum_{i\in\I_0}
 H(C_i|C_1^{i-1},\YY)
 \\
 \Prob(\PPsi(A\XX,\YY)\neq \XX)
 &\leq
 \frac1{\log 2}
 \sum_{i\in\I_0}
 H(C_i|C_1^{i-1},\YY).
\end{align*}
\end{lem}
\begin{IEEEproof}
The first inequality is shown from
Lemmas~\ref{lem:exc}--\ref{lem:error-ssc} as
\begin{align}
 &
 \Prob(\ppsi(A\XX,\YY)\neq \XX)
 \notag
 \\*
 &=
 \Prob(\ff(\CC_1,\YY)\neq (\CC_0,\CC_1))
 \notag
 \\*
 &\leq
 \sum_{i\in\I_0}
 \Prob(f_i(\CC_1^{i-1},\YY)\neq C_i)
 \notag
 \\
 &\leq
 \frac1{2\log 2}
 \sum_{i\in\I_0}
 H(C_i|C_1^{i-1},\YY),
 \label{eq:error-sc}
\end{align}
where the last inequality comes from the relation
\[
 \Prob(\arg\max_u \mu_{U|V}(u|V)\neq U)
 \leq \frac{H(U|V)}{2\log 2}
\]
shown in~\cite{CC66}.
The second inequality is shown similarly as
\begin{align}
 &
 \Prob(\PPsi(A\XX,\YY)\neq\XX)
 \notag
 \\*
 &=
 \Prob(\FF(\CC_1,\YY)\neq (\CC_0,\CC_1))
 \notag
 \\
 &\leq
 \sum_{i\in\I_0}
 \Prob(F_i(\CC_1^{i-1},\YY)\neq C_i)
 \notag
 \\
 &\leq
 2\sum_{i\in\I_0}
 \Prob(f_i(\CC_1^{i-1},\YY)\neq C_i)
 \notag
 \\
 &\leq
 \frac1{\log 2}
 \sum_{i\in\I_0}
 H(C_i|C_1^{i-1},\YY),
 \end{align}
 where the second inequality comes from \cite[Lemma 3]{SDECODING}.
\end{IEEEproof}

The above lemma implies that the error probability of SC/SSC decoding
is small when $\sum_{i\in\I_0}H(C_i|C_1^{i-1},\YY)$ is small.
The following lemma introduces {\em quasi-polarization},
where the both (\ref{eq:polarize0}) and (\ref{eq:polarize1})
are satisfied for all $\delta>0$ and sufficiently large $n$.
It should be noted here that
(\ref{eq:polarize0}) implies that $H(C_i|C_1^{i-1})$ is close to $0$
but (\ref{eq:polarize1}) may not imply
that $H(C_i|C_1^{i-1})$ is close to $1$.
\begin{lem}
The condition
\begin{equation}
 \sum_{i\in\I_0}H(C_i|C_1^{i-1},\YY)\leq \delta
 \label{eq:polarize0}
\end{equation}
is equivalent to the condition
\begin{equation}
 \sum_{i\in\I_1}H(C_i|C_1^{i-1},\YY)\geq H(\XX|\YY)-\delta.
 \label{eq:polarize1}
\end{equation}
\end{lem}
\begin{IEEEproof}
 Since $[A,B]$ is bijective,
 we have the fact that $H(A\XX,B\XX|\YY)=H(\XX|\YY)$.
 Then the condition (\ref{eq:polarize1}) is derived from
 (\ref{eq:polarize0}) as
 \begin{align}
  &\sum_{i\in\I_1}H(C_i|C_1^{i-1},\YY)
  \notag
  \\*
  &=
  \sum_{i=1}^nH(C_i|C_1^{i-1},\YY)
  -\sum_{i\in\I_0}H(C_i|C_1^{i-1},\YY)
  \notag
  \\
  &\geq
  H(\CC_1,\CC_0|\YY)-\delta
  \notag
  \\
  &=
  H(A\XX,B\XX|\YY)-\delta
  \notag
  \\
  &=
  H(\XX|\YY)-\delta,
 \end{align}
 and the condition (\ref{eq:polarize0})
 is derived from (\ref{eq:polarize1}) as
 \begin{align}
  &\sum_{i\in\I_0}H(C_i|C_1^{i-1},\YY)
  \notag
  \\*
  &=
  \sum_{i=1}^nH(C_i|C_1^{i-1},\YY)
  -\sum_{i\in\I_1}H(C_i|C_1^{i-1},\YY)
  \notag
  \\
  &\leq
  H(\CC_1,\CC_0|\YY)-\lrB{H(\XX|\YY)-\delta}
  \notag
  \\
  &=
  H(A\XX,B\XX|\YY)-H(\XX|\YY)+\delta
  \notag
  \\
  &=
  \delta.
 \end{align}
\end{IEEEproof}

The following lemma asserts that
we have the quasi-polarization
when the SC/SSC decoding is effective in the sense that
the binary entropy of the error probability is $o(1/n)$.
\begin{lem}
\label{lem:H<error}
Let $\ppsi$ and $\PPsi$ be the SC and SSC decoding functions,
respectively.
Let $h(\theta)\equiv -\theta\log(\theta)-[1-\theta]\log(1-\theta)$
be the binary entropy function, where the base of $\log$ is $|\X|$.
Then we have
\begin{align*}
 \sum_{i\in\I_0}H(C_i|C^{i-1},Y^n)
 &\leq n\lrB{\e+h(\e)}
 \\
 \sum_{i\in\I_0}H(C_i|C^{i-1},Y^n)
 &\leq n\lrB{E+h(E)},
\end{align*}
where 
\begin{align*}
 \e
 &\equiv\Prob(\ppsi(A\XX,\YY)\neq \XX)
 \\
 E
 &\equiv\Prob(\PPsi(A\XX,\YY)\neq \XX).
\end{align*}
\end{lem}
\begin{IEEEproof}
 In the following, we show the first inequality,
 where we can show the second inequality similarly.
 Let
 \begin{align*}
  \e_i
  &\equiv \Prob(f_i(C_{i-1},\YY)\neq C_i).
 \end{align*}
 Then we have
 \begin{align}
  \sum_{i\in\I_0}H(C_i|C^{i-1},Y^n)
  &\leq
  \sum_{i\in\I_0}
  \lrB{\e_i\log|\X|+h(\e_i)}
  \notag
  \\
  &\leq
  \sum_{i\in\I_0}\lrB{\e\log|\X|+h(\e)}
  \notag
  \\
  &\leq
  n\lrB{\e+h(\e)},
 \end{align}
 where the first inequality comes from the Fano inequality,
 the second inequality comes from the fact that 
 $f_i(c_1^{i-1},\yy)\neq c_i$ implies $\ppsi(A\xx,\yy)\neq \xx$
 and the last inequality comes from the fact that
 $|\I_0|\leq n$.
\end{IEEEproof}

\begin{rem}
We have several interpretations of Lemmas~\ref{lem:error<H}
and~\ref{lem:H<error}.
Lemma~\ref{lem:error<H} asserts that
the SC/SSC decoding is effective when we have the quasi-polarization
and Lemma~\ref{lem:H<error} asserts that
the SC/SSC decoding is not effective when we do not have the
sufficient\footnote{In this statement, `sufficient' means that
 $\sum_{i\in\I_0}H(C_i|C^{i-1},Y^n)=o(n)$.}
quasi-polarization.
Conversely, Lemma~\ref{lem:H<error} asserts that
we have the quasi-polarization
when the SC/SSC decoding is sufficiently effective
and Lemma~\ref{lem:error<H} asserts that
we do not have the quasi-polarization when the SC/SSC decoding is not
effective.
\end{rem}

\begin{rem}
It is mentioned in~\cite[``Polarization is commonplace'']{A11} that
a random permutation of the set $\{0,1\}^n$
is a good polarizer with a high probability.
We can show a similar fact regarding
a good source code $(A,\pphi)$ and a matrix $B$
that introduces extended codewords,
where the index sets $\I_1$ and $\I_0$ are ordered.
We have a slightly tighter bound than Lemma~\ref{lem:H<error}
as follows:
\begin{align}
 \sum_{i=l+1}^nH(C_i|C_1^{i-1},\YY)
 &=
 H(C_{l+1}^n|C_1^l,\YY)
 \notag
 \\
 &=
 H(B\XX|A\XX,\YY)
 \notag
 \\
 &\leq
 H(\XX|A\XX,\YY)
 \notag
 \\
 &\leq
 H(\XX|\pphi(A\XX,\YY))
 \notag
 \\
 &\leq
 \e\log|\X^n|+h(\e)
 \notag
 \\
 &\leq
 \e
 \lrB{n+\log\frac1{\e}+\log e},
 \label{eq:H<error-phi}
\end{align}
where $e$ is the base of the natural logarithm,
the second inequality comes from \cite[Lemma 3.12]{G10},
the third inequality comes from the Fano inequality,
and the fourth inequality comes from the fact that
\begin{align}
 h(\e)
 &=\e\log\frac 1{\e}+[1-\e]\log\lrsb{1+\frac{\e}{1-\e}}
 \notag
 \\
 &\leq
 \e\log\frac 1{\e}+\e\log e
 \notag
 \\
 &=
 \e\lrB{\log\frac 1{\e}+\log e},
\end{align}
by using the relation $\log(1+\theta)\leq\theta\log e$.
This means that we have quasi-polarization when $\e=o(1/n)$.
In particular,
when $\e$ goes to zero exponentially as $n\to\infty$,
$\sum_{i=l+1}^nH(C_i|C_1^{i-1},\YY)$
also goes to zero exponentially.
It should be noted that the combination of Lemma~\ref{lem:error<H}
and (\ref{eq:H<error-phi}) provides bounds slightly different 
from those provided by Theorems~\ref{thm:sc} and~\ref{thm:ssc}.
\end{rem}

It is a future challenge
to find the function $B$ and the order of the index sets
for a general linear code
where $\sum_{i\in\I_0}H(C_i|C_1^{i-1},\YY)$ is small
or $\sum_{i\in\I_1}H(C_i|C_1^{i-1},\YY)$ is close to $H(\XX|\YY)$.
We can expect to reduce the time complexity of SC/SSC decoding
while maintaining sufficient precision of the computation
for the conditional probability distribution $\mu_{C_i|C_1^{i-1}\YY}$.

\section{Stochastic Successive-Cancellation Decoding of Polar Source Code}

In this section, we revisit the polar source code
for a pair $(X,Y)$ of stationary memoryless source
introduced in~\cite{A10,S12}.
For simplicity, we assume that $|\X|$ is a prime number.
For a given positive integer $k$, let $n\equiv 2^k$.
The source polarization transform $G$ is defined as
\[
 G\equiv
 \begin{pmatrix}
  1 & 0
  \\
  1 & 1
 \end{pmatrix}^{\otimes k}
 S_{\mathrm{BR}},
\]
where $\otimes k$ denotes
the $k$-th Kronecker power and $S_{\mathrm{BR}}$ is
the bit-reversal permutation matrix defined in~\cite{A09}.
Then the extended codeword $\cc\in\X^n$ of a source output
$\xx\in\X^n$ is defined as
$\cc\equiv {}^tS_{\mathrm{BR}}{}^tG\xx$,
where both $\cc$ and $\xx$ are column vectors.

From~\cite[Theorem 4.10]{S12}, we have
\begin{equation}
 \lim_{n\to\infty}
 \frac{\lrbar{\lrb{i\in\N: Z(C_i|C_1^{i-1},\YY)\leq 2^{-{n^\beta}}}}}n
 = 1-H(X|Y)
 \label{eq:polarizeZ}
\end{equation}
for all $\beta\in(0,1/2)$, where $Z$ is the source Bhattacharyya
parameter defined as
\[
 Z(U|V)
 \equiv
 \frac 1{|\X|-1}
 \sum_{\substack{
   u,u'\in\X\\
   u\neq u'
 }}
 \sum_v
 \sqrt{\mu_{UV}(u,v)\mu_{UV}(u',v)}.
\]

Let $\I_0$ and $\I_1$ be defined as
\begin{align*}
 \I_0
 &\equiv\lrb{
  i\in\N:
  Z(C_i|C_1^{i-1},\YY)\leq 2^{-{n^\beta}}
 }
 \\
 \I_1
 &\equiv\lrb{
  i\in\N:
  Z(C_i|C_1^{i-1},\YY)> 2^{-{n^\beta}}
  }.
\end{align*}
Then, from (\ref{eq:polarizeZ}),
we have the fact that the encoding rate $|\I_1|/n$ approaches $H(X|Y)$
as
\begin{align}
 \lim_{n\to\infty}\frac{|\I_1|}n
 &=
 \lim_{n\to\infty}\frac{n-|\I_0|}n
 \notag
 \\
 &=
 1-\lim_{n\to\infty}\frac{|\I_0|}n
 \notag
 \\
 &=
 H(X|Y).
\end{align}
Furthermore, from Lemma~\ref{lem:error<H}, we have
\begin{align}
 &
 \lim_{n\to\infty}\Prob(\ppsi(A\XX,\YY)\neq \XX)
 \notag
 \\*
 &\leq
 \lim_{n\to\infty}
 \frac1{2\log 2}
 \sum_{i\in\I_0}
 H(C_i|C_1^{i-1},\YY)
 \notag
 \\
 &\leq
 \frac1{2\log 2}
 \lim_{n\to\infty}
 \sum_{i\in\I_0}
 [|\X|-1]Z(C_i|C_1^{i-1},\YY)
 \notag
 \\
 &\leq
 \frac{|\X|-1}{2\log 2}
 \lim_{n\to\infty}
 n2^{-{n^\beta}}
 \notag
 \\
 &=0
\end{align}
for all $\beta\in(0,1/2)$,
where the second inequality comes from the relation
\[
 H(U|V)\leq \log(1+[|\X|-1]Z(U|V))\leq [|\X|-1]Z(U|V)
\]
shown in~\cite[Eq.~(5)]{A10}, \cite[Eq.~(4.11)]{S12}.
This implies the well-known fact that SC decoding of the polar
source code is effective~\cite{A10,S12}.

Similarly, we have the following theorem,
which implies the effectiveness of the SSC decoding
of the polar source code.

\begin{thm}
\begin{align}
 \lim_{n\to\infty}\Prob(\PPsi(A\XX,\YY)\neq \XX)
 &\leq
 \frac{|\X|-1}{\log 2}
 \lim_{n\to\infty}
 n2^{-{n^\beta}}
 \notag
 \\
 &=0
\end{align}
for all $\beta\in(0,1/2)$.
\end{thm}

\section{Concluding Remarks}

It should be noted that we cannot judge from
Theorems~\ref{thm:smap}--\ref{thm:ssc}
which decoder (SMAP, SC, or SSC) performs the best
when we use the same encoding function $A$.
It is a future challenge to clarify the best decoder theoretically or
empirically.

In Theorems~\ref{thm:smap}--\ref{thm:ssc},
we have assumed that
we can compute the conditional probability distribution defined by
(\ref{eq:sum-product}) exactly.
However, 
the sum-product algorithm may not provide the exact computation of
(\ref{eq:sum-product}).
It is a future challenge
to estimate the approximation error caused by the sum-product algorithm
and to introduce an alternative algorithm
that provides a good approximation.

The following comments on the computational complexity
of the decoding algorithms.
When $A$ is a sparse matrix with the maximum row weight $w$
and we use the Fourier transform\footnote{
 When $|\X|$ is a power of a prime $p$,
 the term $|\X|^2$ can be replaced by $|\X|\log_p|\X|$
 by using the Fast-Fourier-Transform.}
to compute the convolutions in the sum-product algorithm,
the computational complexity of SMAP decoding is
$O(\iota lw[|\X|^2+|\X|w])$,
where $\iota$ denotes the number of iterations
of the sum-product algorithm.
The computational complexity of SC/SSC decoding is
$O(\iota [n-l]lw[|\X|^2+|\X|w])$.
The computational complexity of the SC/SSC decoding of the polar
source code is $O(|\X|^2n\log n)$
by using the recursive construction of $G$~\cite[Section 4.4]{S12}.

\appendix

\begin{lem}
\label{lem:mapUgVW}
For any triplet $(U,V,W)$ of random variables, we have
\begin{align*}
 &
 \Prob(\arg\max_u \mu_{U|VW}(u|V,W)\neq U)
 \notag
 \\*
 &\leq
 \Prob(\arg\max_u \mu_{U|V}(u|V)\neq U).
\end{align*}
\end{lem}
\begin{IEEEproof}
 For all $v$, we have
 \begin{align}
  &
  \sum_{w}\mu_{W|V}(w|v)\max_u\mu_{U|VW}(u|v,w)
  \notag
  \\*
  &\geq
  \max_u\sum_{w}\mu_{W|V}(w|v)\mu_{U|VW}(u|v,w)
  \notag
  \\
  &=
  \max_u\mu_{U|V}(u|v).
 \end{align}
 Then we have
 \begin{align}
  &
  \Prob(\arg\max_u \mu_{U|VW}(u|V,W)\neq U)
  \notag
  \\*
  &=
  1-\sum_v\mu_V(v)\sum_w\mu_{W|V}(w|v)
  \max_u \mu_{U|VW}(u|v,w)
  \notag
  \\
  &\leq
  1-\sum_v\mu_V(v)\max_u \mu_{U|V}(u|v)
  \notag
  \\
  &=
  \Prob(\arg\max_u \mu_{U|V}(u|V)\neq U).
 \end{align}
\end{IEEEproof}

\begin{lem}
\label{lem:U=U'}
Assume that a triplet $(U,V,W)$ of random variables
satisfies $U=V$. Then we have
\[
 \mu_{UW}(u,w)=\mu_{VW}(u,w)\quad\text{for all}\  (u,w).
\]
\end{lem}
\begin{IEEEproof}
 Since $U=V$, the joint distribution of $(U,V,W)$ is given as
 \[
  \mu_{UVW}(u,v,w)
  =\mu_{UW}(u,w)\chi(v=u)
  =\mu_{VW}(v,w)\chi(u=v)
 \]
 for each $(u,v,w)$.
 Then we have
 \begin{align}
  \mu_{UW}(u,w)
  &=
  \sum_v\mu_{UW}(u,w)\chi(v=u)
  \notag
  \\
  &=
  \sum_v\mu_{UVW}(u,v,w)
  \notag
  \\
  &=
  \sum_v\mu_{VW}(v,w)\chi(u=v)
  \notag
  \\
  &=
  \mu_{VW}(u,w),
 \end{align}
 where the last equality comes from the fact that
 $\mu_{VW}(v,w)\chi(u=v)=0$ when $v\neq u$.
\end{IEEEproof}

\section*{Acknowledgments}
The author thanks Dr.~S.~Miyake, Prof.~K.~Iwata,
and Dr.~Y.~Sakai for helpful discussions.

\end{document}